\documentclass[prd,aps,nofootinbib]{revtex4}
\usepackage{amsmath,amssymb,dsfont}

\RequirePackage[T1]{fontenc}

\RequirePackage{graphicx}
\RequirePackage{mathptmx}      
\RequirePackage{flushend}

\makeindex
\usepackage{amsmath}
\usepackage{amssymb}
\begin{document}

\title{A chargeless complex vector matter field in supersymmetric scenario} 

\author{L.P. Colatto}
\email{lcolatto@gmail.com}
\affiliation{CEFET-RJ UnED-Petr\'opolis, CEP 25620-003, RJ, Brasil} 

\author{A.L.A. Penna}
\email{penna.andre@gmail.com}
\affiliation{International Center for Condensed Matter Physics, CP 04513, CEP 70919-970, University of Bras\'{i}lia-DF, Brasil}

\begin{abstract}
In this paper we construct and study a formulation of a chargeless complex vector matter field in a supersymmetric framework. To this aim we combine two no-chiral scalar superfields in order to take the vector component field to build the chargeless complex vector superpartner where the respective field strength transforms as matter fields by a global $U(1)$ gauge symmetry. To the aim to deal with consistent terms without breaking the global $U(1)$ symmetry it imposes a choice to the complex combination revealing a kind of symmetry between the choices and eliminate the extra degrees of freedom consistently with the supersymmetry. As the usual case the mass supersymmetric sector contributes as a complement to dynamics of the model.  We obtain the equations of motion of the Proca's type field, for the chiral spinor fields and for the scalar field on the mass-shell which show the same mass as expected. This work establishes the firsts steps to extend the analysis of charged massive vector field in a supersymmetric scenario.
\end{abstract}

%
\maketitle
\section{Introduction}

Matter field dynamics was firstly established by Dirac in a consistent relativistic framework. He has studied the free electron dynamics where its interaction yields the first steps on QED, which was further developed by Feynmann and others \cite{Itzykson,Ramond}. These studies were very important to the formulation and the understanding of QFT, Standard Model and also to the strings theory. Indeed it has been the basement of all theoretical analysis of any dynamics which concerns to integer or half-integer spin particles. In fundamental quantum theory we have classified in two types: boson and fermions respectively. Fermions usually are the constituent of the matter and bosons are the interaction particle \cite{Bogu}. Nevertheless if we are treating to the weak force we have to deal with charged (or not) massive vector (boson) fields which are the intermediate between the protons and neutrons. So we could interpreted as charged vector matter fields.  

Due to the supersymmetry which plays a fundamental role on strings theory fitting together quantum theories of the gravitational interactions, electroweak and strong forces, the studies on supersymmetric theories are of great interest to high energy physicists \cite{wess,nill,ref1,ref2,ref3,ref4,ref5,ref6,ref7,ref8,ref9,ref10,ref11}.
Supersymmetry deals with graded Lie algebra in the unique reliable algebra extension which holds to be consistent with the S-matrix in relativistic quantum field theory
\cite{man,gat,haag,ref12}. Recalling that this special symmetry correlates fermionic and bosonic fields, called superpartners, which puts them together in a superfield formulation. It remarks the important role played by the study of matter-like vector fields to construct appropriated supersymmetric models \cite{zum,ref13}. Moreover supersymmetric models with chiral superfields and global gauge invariance involving matter fields are elegantly constructed \cite{ref11,ref12,ref13}. Thus quarks, leptons and vector bosons which participate in usual gauge theories, as electroweak and chromodynamics, in a supersymmetric extension coexist along their superpartners: squarks, sleptons and the fermionic partner of the vector bosons which a type of vector matter field. For instance, the supersymmetric version of quantum electrodynamics involves a vector supermultiplet whose contents are a massless photon and its spin-$\frac{1}{2}$ superpartner, the photino\cite{ref11,ref12}.

Indeed theoretical formulation of supersymmetric gauge vector field has been largely studied \cite{ref11,ref13,ref14,FerraraGP}. It was shown that gauge vector field component emerges from non-chiral scalar superfields when one uses some suitable constraint (Wess-Zumino) to remove exceeding non-physical components fields \cite{ref12,ref13}. Nevertheless, there is a lack of studies on models that describe supersymmetric vector matter fields. Therefore, one of the aims of this work is the attempt to address this lack in order to further study the interactions which it can be involved. To this purpose we construct a formulation which the vector field $B_{\mu}$ is complex, but chargeless, and massive and it transforms as matter field by the global $U(1)$ group \cite{Bogu}, {\it i.e.}, we would emphasize that $B_{\mu}$ is not a local gauge field but a Proca-type one. Such models are interesting in order to contribute to the understanding of the supersymmetric model of electroweak theory though it contains charged vector particles. Furthermore it can improve our knowledge of the form of nuclear atomic structure and its interaction at high energy. Further, complex vector fields with matter symmetry are relevant to the vacuum polarization theory that can be connected to models  which deals with Lorentz symmetry violation in high energy physics \cite{ref15,colla1,colla2,and1,and2,jac,matter,meu,Bel1,Manoel1,Bel2,Botta,Bel3,Rizzo1,Manoel2,Bel4,Alfaro,Colatto1,Manoel3,Alfaro1,Rizzo2}. Another aim of this work is to obtain the supermultiplet that will accommodates a charged matter vector field and their supersymmetric partners, and also to get the most appropriate supersymmetric action for this field which will be the subject of a forthcoming work. To this purpose we are going to formulate a supersymmetric Lagrangian starting from chargeless non-chiral superfield which contains the vector (matter) field. The present paper is outlined as follows: in Section II we present a model that acommodates two real vector matter field; in Section III we compose the previous model in a complex form and we present the Dirac superspinor field $\Psi$; in section IV we present a general conclusion.

\section{Two chargeless vector matter superfields model}

We are going to present a chargeless (real) formulation for vector matter field.
To this aim we start from a general non-chiral scalar superfield which includes
in the matter multiplet a vector field as irreducible representation of
the Lorentz group. In order to build more ahead a complex extension we
introduce two chargeless real scalar superfields doubling the number of degrees of freedom,
that are written as
\begin{eqnarray}
\label{sucampo1}
\Phi(x^{\mu},\theta_{a},\bar{\theta}_{\dot{a}})&=&C(x)+
\theta^{a}\varphi_{a}(x)+
\bar{\theta}^{\dot{a}}\bar{\varphi}_{\dot{a}}(x)+\theta^{2}[m(x)+i n(x)]+ \bar{\theta}^{2}[m(x)-in(x)]+\theta^{2}\bar{\theta}^{\dot{a}}\left[
\bar{\lambda}_{\dot{a}}(x)-\frac{i}{2}\, {\sigma^{\mu}}_{a \dot{a}}\,
\partial_{\mu} \,\varphi^a(x) \right]+ \nonumber \\
&+&\bar{\theta}^{2}\theta^{a} \left[\lambda_{a}(x)-\frac{i}{2}\,
 {\sigma^{\mu}}_{a \dot{a}}\, \partial_{\mu} \overline{\varphi}^{\dot{a}}
\right] +\theta^{a}\sigma^{\mu}_{a\dot{a}}\bar{\theta}^{\dot{a}}X_{\mu}(x)+
\theta^{2}\bar{\theta}^{2}\left[ D(x)-\frac{1}{4}\Box C(x) \right],
\nonumber\\
\nonumber\\
\label{sucampo2}
\Lambda(x^{\mu},\theta_{a},\bar{\theta}_{\dot{a}})&=& A(x)+
\theta^{a}\chi_{a}(x)+
\bar{\theta}^{\dot{a}}\bar{\chi}_{\dot{a}}(x)+\theta^{2}[\rho(x) + i \, \tau(x)]+ \bar{\theta}^{2}[\rho(x)-i\, \tau(x)]
+ \theta^{2}\bar{\theta}^{\dot{a}}
\left[\bar{\zeta}_{\dot{a}}(x)-\frac{i}{2}\, {\sigma^{\mu}}_{a
\dot{a}}\, \partial_{\mu} \,\chi^a(x) \right]+ \nonumber \\
&+&\bar{\theta}^{2}\theta^{a}\left[\zeta_{a}(x)-\frac{i}{2} \,
{\sigma^{\mu}}_{a \dot{a}}\, \partial_{\mu}
\,\bar{\chi}^{\dot{a}}(x)
 \right]+\theta^{a}\sigma^{\mu}_{a\dot{a}}\bar{\theta}^{\dot{a}}Y_{\mu}(x)+
\theta^{2}\bar{\theta}^{2}\left[S(x)-\frac{1}{4} \, \Box A(x)\right],
\end{eqnarray}
where the superfields $\Phi$ and $\Lambda$ are particular constructions of matter vector
supermultiplet which include real vector fields $X_{\mu}(x)$ and \,$Y_{\mu}(x)$ with
helicity $\pm1$; the fields $\varphi_{a}(x)$, $\lambda_{a}(x)$, $\chi_{a}(x)$ and
$\zeta_{a}(x)$ are two-components Weyl fermions with helicity $\pm\frac{1}{2}$;
and the fields $C(x)$, $D(x)$, $A(x)$, $S(x)$, $M(x)$, $N(x)$, $\rho(x)$, $\tau(x)$ are real
scalar fields with spin-$0$. It is easy to verify that to both superfields the
number of bosonic and fermionic degrees of freedom are the same. We stress that
we only have applied the reality condition on the superfields what does not spoil
the matter structure of these multiplets. Therefore the dynamics to chargeless supersymmetric
vector fields can be obtained through suitable field-strengths which accommodate
the real superfields $\Phi$ and $\Lambda$.

In order to construct the supersymmetric field-strengths for the
real superfields $\Phi$ and $\Lambda$, which we call chargeless
supersymmetric field-strengths, we are going to apply
supersymmetric covariant derivatives on the above scalar
superfields which result in chiral superfields, in such way that
\begin{eqnarray}
W_{a}=-\frac{1}{4}\bar{D}\bar{D}D_{a}\Phi(x^{\mu},\theta_{a},
\bar{\theta}_{\dot{a}}), \;\;\;
\overline{W}_{\dot{a}}=-\frac{1}{4}DD\bar{D}_{\dot{a}}\Phi(x^{\mu},
\theta_{a},
\bar{\theta}_{\dot{a}}),
\end{eqnarray}
and by similarity for $\Lambda$, we have that
\begin{eqnarray}
\Omega_{a}=-\frac{1}{4}\bar{D}\bar{D}D_{a}\Lambda(x^{\mu},\theta_{a},
\bar{\theta}_{\dot{a}}), \;\;\;
\overline{\Omega}_{\dot{a}}=-\frac{1}{4}DD\bar{D}_{\dot{a}}\Lambda(x^{\mu},
\theta_{a},
\bar{\theta}_{\dot{a}}),
\end{eqnarray}
the $W_{a}$ and $\Omega_{a}$ are chiral spinor superfields.

We can redefine the superfields in the chiral superspace coordinates, namely $\Phi(y^{\mu},
\theta_{a})$, $\Phi(z^{\mu},\bar{\theta}_{\dot{a}})$, $\Lambda(y^{\mu},
\theta_{a})$ and \newline $\Lambda(z^{\mu},\bar{\theta}_{\dot{a}})$, such that
$y^{\mu}=x^{\mu}+i\theta\sigma^{\mu}\bar\theta$ and
$z^{\mu}=x^{\mu}-i\theta\sigma^{\mu}\bar\theta\,$. Hence the supersymmetric
covariant derivatives are defined as
\begin{eqnarray}
\label{dersusy2}
D_{a}=\frac{\partial}{\partial\theta_{a}}+
2i\sigma_{a\dot{a}}^{\mu}\bar{\theta}^{\dot{a}}\frac{\partial}{\partial
y^{\mu}}\; \;\;\mbox{and} \;\;\;
\bar{D}_{\dot{a}}=-\frac{\partial}{\partial\bar{\theta}_{\dot{a}}}-
2i\theta^{a}\sigma_{a\dot{a}}^{\mu}\frac{\partial}{\partial
z^{\mu}}.
\end{eqnarray}
According to this definitions we can compute the field-strengths $W_{a}$ and $\Omega_{a}$, and
we have
\begin{eqnarray}
W_{a}(y\,,\,\theta)&=&\lambda_{a}(y)+2\theta_{a}D(y)+
(\sigma^{\mu\nu}\theta)_{a}X_{\mu\nu}(y) -i\theta^{2}\sigma^{\mu}_{a\dot{b}}
\partial_{\mu}\bar{\lambda}^{\dot{b}}(y),\nonumber\\
\nonumber\\
\Omega_{a}(y\,,\,\theta)&=&\zeta_{a}(y)+2\theta_{a}S(y)+
(\sigma^{\mu\nu}\theta)_{a}Y_{\mu\nu}(y)-i\theta^{2}\sigma^{\mu}_{a\dot{b}}\partial_{\mu}\bar{\zeta}^{\dot{b}}(y),
\end{eqnarray}
and similarly for the $\overline{W}_{\dot{a}}$ and
$\overline{\Omega}_{\dot{a}}$. So we are in conditions to
construct the supersymmetric model in terms of the superfields
$W_{a}$ and $\Omega_{a}$ where chargeless vector matter field is present. The kinetic
 part can be written as
\begin{eqnarray}
\label{kinS}
S_{kin}=\int d^4x \, d^2 \theta \, d^2 \bar{\theta}\;
\left\{ W_{a}W^{a}+\overline{W}_{\dot{a}}\overline{W}^{\dot{a}}  +  \Omega_{a}\Omega^{a}+
\overline{\Omega}_{\dot{a}}\overline{\Omega}^{\dot{a}}\right\}.
\end{eqnarray}
We have adopted the usual conventions for the spinor algebra, for
the superspace parametrization, and the translation invariance of
the integral in the chiral coordinates \cite{ref11,ref13}. Then we
obtain that the expression (\ref{kinS}) has the following
component expansion
\begin{eqnarray}
\label{acin1}
S_{kin}&=&\int d^4x \,
\Big\{ -X_{\mu\nu}(x)X^{\mu\nu}(x)-Y_{\mu\nu}(x)Y^{\mu\nu}(x) -4i\lambda^{a}(x)\sigma^{\mu}_{a\dot{b}}\partial_{\mu}
\bar{\lambda}^{\dot{b}}(x)-4i\zeta^{a}(x)\sigma^{\mu}_{a\dot{b}}
\partial_{\mu}\bar{\zeta}^{\dot{b}(x)}+8D^{2}(x)+8S^2(x)\Big\}.
\end{eqnarray}
The action (\ref{acin1}) describes the kinetic part of supersymmetric chargeless vector
field. However, to write the full action that corresponds to the underlined field
theory we can also consider the supersymmetric mass term, given by
\begin{eqnarray}
\label{realm}
S_m &=& \int d^4x\,\,\, \alpha^2 \left[ \Phi^2+\Lambda^2 \right] \nonumber\\
    &=&\int d^4x \,\,\, \alpha^2 \Big[C(x)D(x)-\frac{1}{4}C(x)\Box C(x)+
    \varphi_{a}(x)\lambda^{a}(x)+ \bar{\varphi}_{\dot{a}}(x)\bar{\lambda}^{\dot{a}}(x) -i\bar{\varphi}^{\dot{a}}\bar{\sigma}^{\mu}_{\dot{a}b}\partial_{\mu}\varphi^{b}
+M(x)^{2}+N(x)^{2}+ \nonumber \\
&&+\frac{1}{4}X^{\mu}X_{\mu}+\frac{1}{4}Y^{\mu}Y_{\mu}+A(x)S(x)-\frac{1}{4}A(x)\Box A(x) + \chi_{a}(x) \zeta^{a}(x) +
\bar{\chi}_{\dot{a}}(x) \bar{\zeta}^{\dot{a}}(x)
-i \bar{\chi}^{\dot{a}} \bar{\sigma}^{\mu}_{\dot{a}b} \partial_{\mu}\chi^{b} +\rho(x)^{2}+\tau(x)^{2} \Big],
\end{eqnarray}
where  $\alpha^2$ is the mass parameter. As usual, the ``mass''
part of the action presents kinetic terms, beyond the usual mass
terms, which were eliminated by spinor chirality property of
superfields $W_{a}$ and $\Omega_{a}$. Furthermore, we could infer
that the mass term in the action (\ref{realm}) arises as a
dynamical complement to the supersymmetric vector matter fields.
Indeed supersymmetric matter-like fields are formulated with
chiral spinor superfields.

\section{The chargeless complex vector matter superfield model}

We know that the supersymmetric action for two free vector matter
fields might be built through non-chiral scalar superfields
\cite{ref13}. Moreover, we can see that the degrees of freedom of
this model are compatible with the dynamical free fields in
complex space $\mathbb{C}$. In this section our aim is to derive
the appropriated complex superfield to describe supersymmetric
complex vector fields. To this aim we need strongly define two
complex non-chiral scalar superfields, defined as
\begin{eqnarray}
\label{sucampoo1}
\Sigma(x_{\mu},\theta_{a},\bar{\theta}_{\dot{a}})&=&k(x)+\theta^{a}\xi_{a}(x)+
\bar{\theta}^{\dot{a}}\bar{C}_{\dot{a}}(x)+ \, \theta^{2}l(x)+\bar{\theta}^{2}f(x)+ \theta^{2}\bar{\theta}^{\dot{a}}\left[
\bar{G}_{\dot{a}}(x)-\frac{i}{2}\, {\sigma^{\mu}}_{a \dot{a}}\,
\partial_{\mu} \,\xi^a(x) \right] + \nonumber \\
&+&\bar{\theta}^{2}\theta^{a} \left[R_{a}(x)-\frac{i}{2}\, {\sigma^{\mu}}_{a \dot{a}}\, \partial_{\mu} \overline{C}^{\dot{a}} \right] +\theta^{a}\sigma^{\mu}_{a\dot{a}}\bar{\theta}^{\dot{a}}B_{\mu}(x)+
\theta^{2}\bar{\theta}^{2}\left[ d(x)-\frac{1}{4}\Box k(x) \right].
\end{eqnarray}
\begin{eqnarray}
\label{sucampoo2}
K(x_{\mu},\theta_{a},\bar{\theta}_{\dot{a}})&=&a(x)+\theta^{a}T_{a}(x)+
\bar{\theta}^{\dot{a}}\bar{H}_{\dot{a}}(x)+\, \theta^{2}j(x)+ \bar{\theta}^{2}E(x)+\theta^{2}\bar{\theta}^{\dot{a}}
\left[\bar{Q}_{\dot{a}}(x)-\frac{i}{2}\, {\sigma^{\mu}}_{a
\dot{a}}\, \partial_{\mu} \,T^a(x) \right]+ \nonumber \\
&+&\bar{\theta}^{2}\theta^{a}\left[U_{a}(x)-\frac{i}{2} \, {\sigma^{\mu}}_{a \dot{a}}\, \partial_{\mu} \,\bar{H}^{\dot{a}}(x) \right]+\theta^{a}\sigma^{\mu}_{a\dot{a}}\bar{\theta}^{\dot{a}}Z_{\mu}(x)+
\theta^{2}\bar{\theta}^{2}\left[v(x)-\frac{1}{4} \, \Box
a(x)\right],
\end{eqnarray}
and similarly for their complex conjugated superfields.

The superfields (\ref{sucampoo1}) and (\ref{sucampoo2}) present multiplets with complex vector fields $B_{\mu}(x)$ e $Z_{\mu}(x)$ and spin-$1$; the $\xi_{a}(x)$, $\bar{C}_{\dot{a}}(x)$, $R_{a}(x)$, $\bar{G}_{\dot{a}}(x)$, $T_{a}(x)$,
$\bar{H}_{\dot{a}}(x)$, $U_{a}(x)$ and $\bar{Q}_{\dot{a}}(x)$ are Weyl fermion fields with spin-$\frac{1}{2}$; and the $k(x)$, $d(x)$, $a(x)$ and $v(x)$ are complex scalar fields with spin-$0$. So, to construct the complex extension for the chargeless vector matter fields we combine the superfields $\Phi$ and $\Lambda$ in such way that
\begin{eqnarray}
\label{defcomplexas}
\Sigma=\Phi+i\Lambda \;\;\; \mbox{or} \;\;\; K=\Lambda+i\Phi.
\end{eqnarray}
The combinations (\ref{defcomplexas}) are the two possible realizations of complex extension of chargeless vector matter fields. And it is clear that they have relation via complex algebra. 
Indeed the transformation rule that implies in an invariant choice mechanism is given by
\begin{eqnarray}
\label{eqval}
K=i\Sigma^{\dag}.
\end{eqnarray}

This means that one can choose freely between the two equivalent complex extensions in (\ref{defcomplexas}) 
with no loss of generality of dynamical structure of the fields. We observe that the transformation rule  (\ref{eqval}) 
guarantees to write
a consistent kinetic term for the complex vector field without the breaking of the global $U(1)$ gauge symmetry. Another advantage that came to light is that the transformations (\ref{eqval}) 
eliminate the exceeding fields which does not contribute for the supersymmetric action, what allow bosons and fermions to have the same physical degrees of freedom. Indeed the constraint relation to the superfields imply to the relations of the component fields as follows
\begin{eqnarray}
\label{esp1}
a(x)&=& ik^{*}(x),\; T_{a}(x)= iC_{a}(x),\; v(x)=id^{*}(x) \nonumber \\
\bar{H}_{\do{a}}(x)&=& i\bar{\xi}_{\do{a}}(x),\; j(x)= if^{*}(x),\;U_{a}(x)= iG_{a}(x), \nonumber \\
E(x)&=& il^{*}(x),\; \bar{Q}_{\do{a}}(x)= i\bar{R}_{\do{a}}(x),\;Z_{\mu}(x)= iB^{*}_{\mu}(x).
\end{eqnarray}
So, we can adjust the complex extension of the chargeless superfields $\Phi$ and $\Lambda$ by assuming the equation $\Sigma=\Phi+i\Lambda$, where we find the following relation of fields
\begin{eqnarray}
\label{idecampos1}
k(x)&=& C(x)+iA(x),\;\xi_{a}(x)=\varphi_{a}(x)+i\chi_{a}(x),\nonumber\\
\bar{C}_{\do{a}}(x)&=& \bar{\varphi}_{\do{a}}(x)+i\bar{\chi}_{\do{a}}(x),\nonumber \\
l(x)&=& m(x)+n(x)+i(\rho(x)+\tau(x)),\nonumber\\
\bar{G}_{\do{a}}(x)&=& \bar{\lambda}_{\do{a}}(x)+i\bar{\zeta}_{\do{a}}(x),\nonumber \\
f(x)&=& m(x)+ n(x)-i(\rho(x)+\tau(x)),\nonumber\\
R_{a}(x)&=& \lambda_{a}(x)+i\zeta_{a}(x),\;B_{\mu}(x)= X_{\mu}(x)+iY_{\mu}(x),\nonumber\\
d(x)&=& D(x)+iS(x).
\end{eqnarray}

To describe the dynamics of the supersymmetric complex vector fields with matter symmetry we need to construct an appropriated complex supersymmetric field-strength model in order to
accommodate the superfields $\Sigma$ and $K$. This can be reached starting from the following definitions
\begin{eqnarray}
\label{chargedfs}
\Upsilon_{a}&=&-\frac{1}{4}\bar{D}\bar{D}D_{a}\Sigma(x_{\mu},\theta_{a},\bar{\theta}_{\dot{a}}), \nonumber \\
\bar{\Upsilon}_{\dot{a}}&=&-\frac{1}{4}DD\bar{D}_{\dot{a}}\Sigma^{\dag}(x_{\mu},\theta_{a},
\bar{\theta}_{\dot{a}}),\nonumber\\
\Gamma_{a}&=&-\frac{1}{4}\bar{D}\bar{D}D_{a}K(x_{\mu},\theta_{a},\bar{\theta}_{\dot{a}}),\nonumber \\
\bar{\Gamma}_{\dot{a}}&=&-\frac{1}{4}DD\bar{D}_{\dot{a}}K^{\dag}(x_{\mu},\theta_{a},\bar{\theta}_{\dot{a}}),
\end{eqnarray}
where $\Upsilon_{a}$ and $\Gamma_{a}$ are charged spinor superfields. As a consequence of the complex extension procedure we must relate the chargeless spinor superfields $\Omega_{a}$ and $W_{a}$ with the complex definitions (\ref{chargedfs}) which, in the simplest way, is
\begin{eqnarray}
&&\Upsilon_{a}=W_{a}+i\Omega_{a},\hspace{3cm}
\bar{\Upsilon}_{\dot{a}}=\bar{W}_{\dot{a}}-i\bar{\Omega}_{\dot{a}},\nonumber\\
\nonumber\\
&&\Gamma_{a}=\Omega_{a}+iW_{a},
\hspace{3cm}\bar{\Gamma}_{\dot{a}}=\bar{\Omega}_{\dot{a}}-i\bar{W}_{\dot{a}}.
\end{eqnarray}
and by assuming the spinor identities $W_{a}\Omega^{a}=\Omega_{a}W^{a}$ and $ \bar{W}_{\dot{a}}\bar{\Omega}^{\dot{a}}=\bar{\Omega}_{\dot{a}}\bar{W}^{\dot{a}}$ we can find the kinetic supersymmetric Lagrangian for the complex vector fields
\begin{equation}
\label{lagranC}
{\cal
L}_{k}=i\big(\bar{\Upsilon}_{\dot{a}}\bar{\Gamma}^{\dot{a}}-\Upsilon_{a}\Gamma^{a}\big)=W_{a}W^{a}+\bar{W}_{\dot{a}}\bar{W}^{\dot{a}}+\Omega_{a}\Omega^{a}
+\bar{\Omega}_{\dot{a}}\bar{\Omega}^{\dot{a}}.
\end{equation}

We can observe that the left-hand side of the latter equation is the complex extension of chargeless Lagrangian (\ref{acin1}) that was written in terms of charged spinor superfields. Bearing this in mind, we can then redefine the kinetic Lagrangian (\ref{lagranC}) simply by combining
the charged spinor superfields $\Upsilon_{a}$ and $\Gamma_{a}$ as a ``Dirac superspinor'' $\Psi$, such that
\begin{equation}
\Psi(x^{\mu},\theta_{a},\bar{\theta}_{\dot{a}})= \left(\begin{array}{c} \Upsilon_a  \\ \bar{\Gamma}^{\dot{a}} \end{array} \right),
\end{equation}
and also we assume $\overline{\Psi}$  as the adjoint Dirac superspinor representation.
In this case we have that $\overline{\Psi}=\Psi^{\dag}\gamma^{0}=( \Gamma^{a} \; \bar{\Upsilon}_{\dot{a}})$, and so the supersymmetric action from the kinetic Lagrangian (\ref{lagranC}) is now given by
\begin{eqnarray}
\label{acinv}
S_{k}&=&\int d^4x \, d^2 \theta \, d^2 \bar{\theta}\,\,
\big(i\overline{\Psi}\Psi\big) = \nonumber \\
&=&i\int d^4x \, d^2 \theta \,d^2
\bar{\theta}\,\,\big(\bar{\Upsilon}_{\dot{a}}\bar{\Gamma}^{\dot{a}}-\Upsilon_{a}\Gamma^{a}\big).
\end{eqnarray}
We can note that the product of Dirac superspinors $\overline{\Psi}\Psi$ obeys matter symmetry and it presents an interesting analogy to charged scalar superfield product $S^{\dag}S$. In this sense we verify that $\Psi$ and $\overline{\Psi}$ represent two chiral supersymmetric extensions for the matter vector field which can be transformed under $U(1)$ global gauge group in the follow way
\begin{eqnarray}
\Psi'=e^{-2iq\beta}\Psi\hspace{3.0cm}\overline{\Psi}'&=&\overline{\Psi}e^{2iq\beta^{\dag}},
\label{transf}
\end{eqnarray}
where $\beta$ is a global $U(1)$ gauge parameter, $q$ is the charge of the global symmetry. So the action (\ref{acinv}) is then invariant under the transformations \eqref{transf}. In order to obtain the component Lagrangian we can expand the product $\overline{\Psi}\Psi$ by considering that
\begin{eqnarray}
\Upsilon_{a}(y\,,\,\theta)&=&R_{a}(y)+2\theta_{a}d(y)+(\sigma^{\mu\nu}\theta)_{a}F_{\mu\nu}(y) -i\theta^{2}\sigma^{\mu}_{a\dot{b}}\partial_{\mu}\bar{G}^{\dot{b}}(y) , \nonumber\\
\Gamma_{a}(y\,,\,\theta)&=&U_{a}(y)+2\theta_{a}v(y)+(\sigma^{\mu\nu}\theta)_{a}Z_{\mu\nu}(y) -i\theta^{2}\sigma^{\mu}_{a\dot{b}}\partial_{\mu}\bar{Q}^{\dot{b}}(y),
\end{eqnarray}
and similarly for $\bar{\Upsilon}_{\dot{a}}$ and $\bar{\Gamma}_{\dot{a}}$. 
We note the presence of the complex matter field-strengths, namely
\begin{eqnarray}
F_{\mu\nu}=\partial_{\mu}B_{\nu}-\partial_{\nu}B_{\mu} \;\;\; \mbox{and} \;\;\;
Z_{\mu\nu}=\partial_{\mu}Z_{\nu}-\partial_{\nu}Z_{\mu}
\end{eqnarray}
hence the action (\ref{acinv}) can be expanded and we obtain
\begin{eqnarray}
\label{acinc}
S_{k}=\int d^4x &
\Big\{&\frac{i}{2}F_{\mu\nu}(x)Z^{\mu\nu}(x)-\frac{i}{2}{F^{*}}_{\mu\nu}(x){Z^*}^{\mu\nu}(x) -R^{a}(x){\sigma^{\mu}}_{a\dot{b}}\,\partial_{\mu}\bar{Q}^{\dot{b}}(x)-U^{a}(x){\sigma^{\mu}}_{a\dot{b}}\,\partial_{\mu}\bar{G}^{\dot{b}}(x) + \nonumber \\
&+&\bar{R}^{\dot{a}}(x){{\overline{\sigma}^{\mu}}}_{\dot{a}b}\,\partial_{\mu}Q^{b}(x)
+\bar{U}^{\dot{a}}(x){\overline{\sigma}^{\mu}}_{\dot{a}b}\,\partial_{\mu}G^{b}(x)+4i\,v(x)\,d(x)-4i\,v^{*}(x)\,d^{*}(x)\Big\}.
\end{eqnarray}
In this format we can recognize the dynamical term that describes the matter vector field as
$\frac{i}{2}F_{\mu\nu}Z^{\mu\nu}-\frac{i}{2}F^{*}_{\mu\nu}{Z^*}^{\mu\nu}$. It involves both $F_{\mu\nu}$ and $Z_{\mu\nu}$ matter tensors. However, it does not correspond to the conventional kinetic term for the matter vector field, and the action (\ref{acinv}) shows more degrees of freedom than it is necessary. In order to get rid of
such fields we must assume the rule of transformation (\ref{eqval}) which is a constraint of half of the degrees and consequently the action (\ref{acinv}) reach the correct number of component fields.

Applying the condition (\ref{eqval}) in the action (\ref{acinc}) we can reach the usual dynamical matter field strength term, or
\begin{eqnarray}
\frac{i}{2}F_{\mu\nu}Z^{\mu\nu}-\frac{i}{2}{F^{*}}_{\mu\nu}{Z^*}^{\mu\nu}
\;\;\Longrightarrow \;\;-{F^{*}}_{\mu\nu}F^{\mu\nu},
\end{eqnarray}
and so the $Z^{\mu\nu}$ tensor field is reabsorbed in this action. Likewise, and without loss of generality, we could have chosen the inverse relation $\Sigma=iK^{\dag}$ what implies to reabsorb the $F^{\mu\nu}$ tensor field. Then by using the whole relation (\ref{esp1}) in action (\ref{acinv}) we find the complex supersymmetric model for the matter vector field can be written as
\begin{eqnarray}
\label{açãofin}
S_{k}&=&\int d^4x \,
\big\{-{F^{*}}_{\mu\nu}(x)F^{\mu\nu}(x)
-2iR^{a}(x){\sigma^{\mu}}_{a\dot{b}}\,\partial_{\mu}\bar{R}^{\dot{b}}(x) -2iG^{a}(x){\sigma^{\mu}}_{a\dot{b}}\,\partial_{\mu}\bar{G}^{\dot{b}}(x)+8d^{*}(x)\,d(x)\big\},
\end{eqnarray}
where the expression $-{F^{*}}_{\mu\nu}(x)F^{\mu\nu}(x)$ represents the usual kinetic term of the vector matter field while the terms represent with the components $R^{a}$ and $G^{a}$ the fermionic sector, and the last term corresponds to the auxiliary field $d$ term.
To completeness we are going to introduce the massive action term in the model. Observing the symmetries of non-chiral fields
$\Sigma$ and $K$ the massive supersymmetric term can be suitable defined as
\begin{eqnarray}
\label{acinM}
    S_{m} =\frac{\alpha^2}{2}\int d^4x \, d^2 \theta \, d^2 \bar{\theta}\; \left[ \Sigma^{\dag}\Sigma
    + K^{\dag}K \right] ,
\end{eqnarray}
where $\alpha^2$ is a mass parameter. From non-chiral superfields $\Sigma$ and $K$ we can obtain the massive vector matter field term ${B^*}_{\mu}B^{\mu}$ as well as their supersymmetric parters. In order to performed it we are going to compute the action (\ref{acinM}) by employing the condition $K=i\Sigma^{\dag}$ where one have that $\frac{1}{2}\Sigma^{\dag}\Sigma+\frac{1}{2}K^{\dag}K=\Sigma^{\dag}\Sigma$, and by applying the definition (\ref{sucampoo2}) the full supersymmetric matter vector field model
can be then obtained from Dirac superspinor field $\Psi$ associated to the non-chiral scalar fields $\Sigma$ and $K$ in the following form
\begin{eqnarray}
\label{acint}
S= S_k+S_m&=&\int d^4x \, d^2 \theta \, d^2 \bar{\theta}\,\,
\big(i\overline{\Psi}\Psi+\alpha^2 \Sigma^{\dag}\Sigma\big) ,
\end{eqnarray}
where the mass part of action can be obtained in component fields as
\begin{eqnarray}
\label{coplm}
    S_{m}&=&\alpha^2 \int d^4x \left(d^{*}(x)k(x)+d(x)k^{*}(x)-\frac{1}{4}k^{*}(x)\Box k(x) -\frac{1}{4}k(x)\Box k^{*}(x)+\xi_{a}(x)G^{a}(x)+\bar{\xi}_{\dot{a}}(x)\bar{G}^{\dot{a}}(x)+ \right. \nonumber \\
&+& C_{a}(x)R^{a}(x)+\bar{C}_{\dot{a}}(x)\bar{R}^{\dot{a}}(x)
-2i\bar{\xi}^{\dot{a}}(x){\overline{\sigma}^{\mu}}_{\dot{a}b}\partial_{\mu}\xi^{b}(x)- 2i\bar{C}^{\dot{a}}(x){\overline{\sigma}^{\mu}}_{\dot{a}b}\partial_{\mu}C^{b}(x)
+{B^{*}}_{\mu}B^{\mu} + f^{*}(x)\,f(x)+l^{*}(x)\,l(x)\Big).
\end{eqnarray}
Here we observe the mass term to $B^{\mu}$,  $f(x)$ and $l(x)$ fields. 
As in the usual supersymmetric models we note that the mass action (\ref{coplm}) also  contributes to kinetic
structure, namely  with the terms $-\frac{1}{4}k^{*}(x)\Box k(x)$, $-2i\bar{\xi}^{\dot{a}}{\overline{\sigma}^{\mu}}_{\dot{a}b}\partial_{\mu}\xi^{b}$, \newline
$-2i\bar{C}^{\dot{a}}{\overline{\sigma}^{\mu}}_{\dot{a}b}\partial_{\mu}C^{b}$. The action also shows mixing mass scalar and fermionic terms, namely  $d^{*}(x)k(x)$, $d(x)k^{*}(x)$,   $\xi_{a}(x)G^{a}(x)$, $\bar{\xi}_{\dot{a}}(x)\bar{G}^{\dot{a}}(x)$, $C_{a}(x)R^{a}(x)$ and $\bar{C}_{\dot{a}}(x)\bar{R}^{\dot{a}}(x)$. By verifying the presence of extra kinetic terms in (\ref{coplm}) it can suggest that when we particularly treat supersymmetric matter vector fields the mass action contributes with a ``dynamic complement'' to the kinetic action
(\ref{açãofin}). Furthermore, we remark that the mass action (\ref{coplm}) is important to match the number of bosonic and fermionic degrees of freedom of the supersymmetric matter vector action (\ref{acint}) for the consistency of the model. We can redefine the component fields absorbing the mass parameter as follows

\begin{eqnarray}
k(x)&\rightarrow&\frac{1}{\alpha}k(x),\;\;\;\xi_{a}(x)\rightarrow\frac{1}{\alpha}\xi_a(x),\;\;\;
C_{a}(x)\rightarrow\frac{1}{\alpha}C_a(x), \nonumber \\
f(x)&\rightarrow&\frac{1}{\alpha}f(x),\;\;\;
l(x)\rightarrow\frac{1}{\alpha}l(x).
\end{eqnarray}

So the action can be rewritten as
\begin{eqnarray}
\label{açãofin2}
S_{ek}&=&\frac{1}{2}\int d^4x \,
\Big(-{F^{*}}_{\mu\nu}(x)F^{\mu\nu}(x)-2iR^{a}(x){\sigma^{\mu}}_{a\dot{b}}\partial_{\mu}\bar{R}^{\dot{b}}(x) -2iG^{a}(x){\sigma^{\mu}}_{a\dot{b}}\partial_{\mu}\bar{G}^{\dot{b}}(x)-\frac{1}{4}k^{*}(x)\Box k(x)-\frac{1}{4}k(x)\Box k^{*}(x)+ \nonumber \\
&-&2i\bar{\xi}^{\dot{a}}(x){\overline{\sigma}^{\mu}}_{\dot{a}b}\partial_{\mu}\xi^{b}(x)-2i\bar{C}^{\dot{a}}(x){\overline{\sigma}^{\mu}}_{\dot{a}b}\partial_{\mu}C^{b}(x)+8d^{*}(x)d(x)+f^{*}(x)f(x)+l^{*}(x)l(x)\Big),
\end{eqnarray}
analogously the mass action is now given by
\begin{eqnarray}
\label{açãofin3}
S_{em}=\int d^4x \,
\Big\{\alpha^2{B^{*}}_{\mu}B^{\mu}+\alpha d^{*}(x)k(x)+\alpha d(x)k^{*}(x)+\alpha\xi_{a}(x)G^{a}(x)+\alpha\bar{\xi}_{\dot{a}}(x)\bar{G}^{\dot{a}}(x)
+\alpha
C_{a}(x)R^{a}(x)+\alpha\bar{C}_{\dot{a}}(x)\bar{R}^{\dot{a}}(x)\Big\},
\end{eqnarray}
so the complex scalar fields $d(x)$, $f(x)$ and $l(x)$ have no
dynamics and arise as auxiliary fields.  Thus assuming the
action (\ref{acint}) as the sum of the redefined actions
(\ref{açãofin2}) and (\ref{açãofin3}) and rearranging\footnote{We
adopt the Weyl representation to the gamma matrices.} the (Dirac)
spinor fields $\Theta$ and $\Pi$ it results in  an off-shell
action $S_{os}$ written in the following form
\begin{eqnarray}
\label{global}
S_{os}=\int d^4x &
\Big\{&-\frac{1}{2}{F^{*}}_{\mu\nu}(x)F^{\mu\nu}(x)+\frac{1}{4}\partial_{\mu}k^{*}(x)\partial^{\mu}k(x) -i\overline{\Theta}(x){\gamma}^{\mu}\partial_{\mu}\Theta(x)
-i\overline{\Pi}(x){\gamma}^{\mu}\partial_{\mu}\Pi(x)
+ \alpha\overline{\Theta}(x){\gamma}^{5}\Theta(x)+ \alpha\overline{\Pi}(x){\gamma}^{5}\Pi(x) + \nonumber \\
&+&4d^{*}(x)d(x)+\frac{1}{2}f^{*}(x)f(x)+\frac{1}{2}l^{*}(x)l(x)+ \alpha^2 B^{*}_{\mu}B^{\mu}+\alpha
d^{*}(x)k(x)+\alpha d(x)k^{*}(x)\Big\},
\end{eqnarray}
where we denote the Dirac spinors of mass $\alpha$ as
\begin{equation}
\Theta(x)=\left(\begin{array}{c}
\xi_{a}\\ \bar{G}^{\dot{a}}\end{array}\right), \;\; \Pi(x)=\left(\begin{array}{c}
C_{a} \\ \bar{R}^{\dot{a}}\end{array} \right).
\end{equation}
For the action (\ref{global}) we have obtained chiral spinor mass
terms given by $\alpha\overline{\Theta}(x){\gamma}^{5}\Theta(x)$
and $\alpha\overline{\Pi}(x){\gamma}^{5}\Pi(x)$. It results that
the motion equations for the fields are
\begin{eqnarray}
&&\partial_{\mu}F^{\mu\nu}(x)+\alpha^2 B^{\nu}(x)=0\nonumber\\
\nonumber\\
&&{\gamma}^{\mu}\partial_{\mu}\Theta(x)-\alpha{\gamma}^{5}\Theta(x)=0,\nonumber\\
\nonumber\\
&&{\gamma}^{\mu}\partial_{\mu}\Pi(x)-\alpha{\gamma}^{5}\Pi(x)=0,\nonumber\\
\nonumber\\
&&\Box k(x)+\alpha^2 k(x)=0,
\label{eqmotion}
\end{eqnarray}

with $f(x)=l(x)=0$. Taking the off-shell action (\ref{global}) we note that it has
$16$-bosonic degrees of freedom concerning to the matter fields
$B_{\mu}(x)$, $k(x)$, $d(x)$, $f(x)$, $l(x)$ and their complex
conjugated ones, as well as $16$-fermionic degrees of freedom for
the Dirac spinor fields $\Theta(x)$ and $\Pi(x)$ and their
conjugated complex ones, what is consistent to the
 supersymmetry. From the equations of motion \eqref{eqmotion} we note that there are three
 auxiliary complex scalar fields $d(x)$, $f(x)$, $l(x)$ and a massive dynamical complex
 scalar field $k(x)$. Moreover, as expected we have obtained a matter Proca-type equation
 for the field $B_{\mu}$. In this context, it is interesting to note that from the
 on-shell action that we can easily extract from the action (\ref{global}) the
 supersymmetric generalization of matter vector field it is only possible if we
 include two dynamical Dirac chiral spinor fields $\Theta(x)$ and $\Pi(x)$ along
 a massive scalar field $k(x)$. Furthermore, a peculiar aspect of the spinor fields
 in the present case is that their mass terms arise as a result of its chiral structure.

\section{Conclusion}

In this paper, which is part of a program started in a previous work \cite{matter}, we propose to formulate and analyze the simple dynamics of chargeless vector matter field in a supersymmetric scenario. There are three motivation for this proposal: the lack of this approach in the literature; to get a clue on the possibility of Lorentz symmetry violation in supersymmetric theories; and the role played by the simplest case of high-spin field in field and string theories. To these aims we have started from real non-chiral scalar superfield in order to obtain real matter Proca-type field in a supersymmetric Lagrangian generalization \cite{matter}. In a straightforward way, the complex model was obtained extending the real scalar non-chiral superfields to the complex space\cite{HelayelNeto1,Kobayashi1,Grassi1,Kuzenko,Gates1}. 

The very interesting point in this enterprise is that in order to obtain a complex Proca-type term, we face to an ambiguous choice of which is the real or imaginary part of the field. This ambiguity reveals to be a symmetry which we represent as a kind of simple Hodge-duality symmetry, $K=i\Sigma^{\dag}$. In fact, despite its apparent simplicity, it is essential to match the number of degrees of freedom of the fermionic and bosonic sectors and to compose the complex (chargeless) Proca-Type dynamical term with global $U(1)$ symmetry. As this symmetry appears by construction, it represents a different approach compared to the Ferrara, Girardello, Palumbo's work \cite{FerraraGP}. On the other hand, as usual, the supersymmetric mass term is also important to compose the dynamics of the supersymmetric model and of its component fields.

We can conclude that in addition to the supersymmetric mass term, to build supersymmetric matter chargeless vector field Lagrangian the Hodge-type duality symmetry is an essential ingredient to be further explored. We can also observe that, as the usual superQED model, the complex dynamical Proca-type term emerges from a product of Dirac superspinors $\Psi\overline{\Psi}$. We can remark that the Dirac superspinor field $\Psi$ is also chiral because it is a combination of chiral Weyl superspinors $\Gamma^{a}$ and $\Upsilon^{a}$.  Finally, we emphasize that the on-shell supersymmetric action obtained (\ref{global}) reveals two fermionic Dirac fields $\Theta(x)$ and $\Pi(x)$, and a massive scalar field $k(x)$ as supersymmetric partners associated to the complex vector field $B_{\mu}(x)$, which is the initial step to implement local gauge symmetry, charge and interaction, we could anticipate that non-minimal couplings have an essential role, in a forthcoming work. 

\section{Acknowledgments}

\noindent
The authors wish to thank Jos\'{e} Hel\"ayel-Neto for valuable discussions.

\end{document}